\documentclass[11pt]{article}

\usepackage[utf8]{inputenc}  
\usepackage[T1]{fontenc}

\usepackage{lmodern}
\usepackage{microtype}
\usepackage{amsmath,amssymb}
\usepackage{mathrsfs}
\usepackage[table]{xcolor} 
\usepackage{graphicx}
\usepackage{braket}
\usepackage{mleftright} 
\usepackage{array}
\usepackage{booktabs}
\usepackage{slashed}
\usepackage{color}   

\usepackage[english]{babel} 
\usepackage{caption} 
\usepackage{subcaption} 
\usepackage{bm} 
\usepackage{multirow} 
\usepackage{calc} 
\usepackage{makecell} 
\usepackage{physics}
\usepackage{url}
\usepackage{hyperref}
\usepackage{cleveref}
\usepackage{ragged2e}     

\definecolor{blizzardblue}{rgb}{0.93, 0.93, 0.93} 
\newcolumntype{?}{!{\vrule width 0.8pt}} 
\mathchardef\mhyphen="2D 

\newcommand{\RN}[1]{ 
  \textup{\uppercase\expandafter{\romannumeral#1}}%
\renewcommand\thesubfigure{(\alph{subfigure})} 
\captionsetup[sub]{labelformat=simple} 
}

\begin{document}

\begin{center}

\vspace*{15mm}
\vspace{1cm}
{\Large \bf New Collider Searches \\\vspace{.2cm} for Axion-like Particles Coupling to Gluons}

\vspace{1cm}

\small{\bf Gholamhossein Haghighat$^{\dagger}$\footnote{h.haghighat@ipm.ir}, Daruosh Haji Raissi$^{\dagger}$\footnote{daruoshraissi96@ipm.ir}, Mojtaba Mohammadi Najafabadi$^{\dagger}$\footnote{mojtaba@ipm.ir}}

 \vspace*{0.5cm}
 
{\small\sl 
$^{\dagger}$ School of Particles and Accelerators, Institute for Research in Fundamental Sciences (IPM), \\P.O. Box 19395-5531, Tehran, Iran}

\vspace*{.2cm}
\end{center}

\vspace*{10mm}

\begin{abstract}\label{abstract}
Axion-like particles (ALPs) are pseudo Nambu-Goldstone bosons associated with spontaneously broken global symmetries emerging in many extensions of the Standard Model. Assuming the most general effective Lagrangian up to dimension-5 operators for an ALP interacting with the SM fields, we investigate for the first time the sensitivity of the LHC13 to the ALP production in association with a di-jet. This study is focused on light ALPs which appear as invisible particles at the detector. Performing a realistic detector simulation and deploying a multivariate technique to best discriminate the signal from backgrounds, we set expected upper bounds on the ALP coupling to gluons. A comprehensive set of background processes is considered, and it is shown that this process provides significant sensitivity to the ALP-gluon coupling and the resulting bound is more stringent than those already obtained at the LHC. We also present prospects for the HE-LHC27 and FCC-hh100 and show that these future colliders are able to improve the limits from the LHC by roughly one order of magnitude.
\end{abstract}

\newpage

\section{Introduction}
\label{sec:introduction}
The Standard Model (SM) of particle physics provides remarkable predictions which have been successfully verified in plenty of experimental observations. However, it suffers from some theoretical and experimental shortcomings. An extensive effort has been therefore made towards the development of theories beyond the Standard Model (BSM) to address some of the open questions. However, no success in observation of these BSMs has been achieved yet.

A long time ago, Peccei and Quinn \cite{Peccei:1977hh} showed in their proposed Axion model that incorporating the anomalous global $U(1)_{\mathrm{PQ}}$ symmetry in the SM Lagrangian can lead to a chiral solution to the strong CP problem~\cite{Dine:2000cj,Hook:2018dlk}. Over time, different aspects of their idea were studied and different variants of their original model were developed~\cite{DiLuzio:2020wdo,Hook:2019qoh,Quevillon:2019zrd,Bellazzini:2017neg,Arganda:2018cuz}. In the Axion model, spontaneous breaking of the anomalous PQ symmetry yields a new particle called the QCD Axion with a mass stringently related to the scale of global symmetry breaking $f_a$. Setting the strict relation between the Axion mass and $f_a$ aside, the idea of Axion-like particles (ALPs) \cite{Brivio:2017ije,Mimasu:2014nea} with wide-ranging masses and couplings arose. Unlike the QCD Axion, the mass and couplings of the ALP are considered as independent parameters. In general, these pseudo Nambu-Goldstone bosons, ALPs, exist in any model with a spontaneously broken global symmetry. They are gauge-singlet, CP-odd and naturally light because of a (softly broken) shift symmetry. 

In recent decades, there has been increasing interest in ALPs due to their various beneficial properties. ALPs display various applications depending on the region in parameter space spanned by their masses and couplings. For instance, they might make good candidates for non-thermal Dark Matter (DM) \cite{Preskill:1982cy,Abbott:1982af,Dine:1982ah}. They can play an essential role in baryogenesis providing a possible explanation for the observed imbalance of matter and antimatter \cite{Jeong:2018jqe,Co:2019wyp}. ALPs might help solve the neutrino mass problem by an ALP-neutrino connection (coupling) through which neutrinos acquire mass \cite{Dias:2014osa, Chen:2012baa, Salvio:2015cja}. They can be used to explain the anomalous magnetic dipole moment of the muon \cite{Chang:2000ii}. The excess events observed in the rare $K$ mesons searches recently reported by the KOTO experiment can be explained by ALPs \cite{Kitahara:2019lws}. Furthermore, ALPs can explain the anomalous decays of the excited Beryllium $^{8}\mathrm{Be}^*$ producing an excess of electron-positron pairs which has been recently reported \cite{Krasznahorkay:2015iga,Ellwanger:2016wfe}. 

Numerous applications that ALPs found to concrete problems provide the motivation to devote considerable attention to these new degrees of freedom. A significant region of the parameter space in terms of the ALP mass and its couplings has been already explored or will be accessible by low-energy experiments, cosmological observations, collider searches, etc.~\cite{Alonso-Alvarez:2018irt,Aloni:2019ruo,Ebadi:2019gij,Braine:2019fqb,Inan:2020aal,Beacham:2019nyx}. ALPs favor derivative couplings to SM particles due to an (approximate) shift symmetry, and the strength of their couplings is proportional to the inverse of global symmetry breaking scale. ALPs with masses below the electron pair production threshold can only decay into a pair of photons, with a decay rate varying as the third power of the ALP mass. Heavier ALPs can also experience the decays into charged leptons and hadrons depending on their masses and couplings. For light ALPs, the decay rate is usually so small that the ALP can travel a long distance before it decays. This motivates searches for solar ALPs (produced by the Primakoff process) using helioscopes such as SUMICO \cite{Inoue:2008zp} and CAST \cite{Arik:2008mq}. The future helioscope experiment IAXO which is a proposed 4th generation axion helioscope is sensitive to ALPs with masses below the GeV scale \cite{Irastorza:2013dav}. In collider searches, long-lived ALPs appear as invisible particles and manifest themselves as missing energy since they decay outside the detector and don't interact with the detecting devices. Searching for missing energy signals, collider experiments probe long-lived ALPs. It has been shown that mono-$\gamma$ and mono-jet plus missing energy are promising signals to probe ALP masses above the KeV scale \cite{Mimasu:2014nea}. Another study has reported mono-W, mono-Z and W+$\gamma$ plus missing energy as promising signals to probe the mass of the ALP and its couplings to electroweak gauge fields \cite{Brivio:2017ije}. Besides the missing energy signals, signals from ALPs decaying inside the detector have also been explored extensively. Searches for the exotic Higgs decays $h\rightarrow aZ$ and $h\rightarrow aa$ with subsequent ALP decays $a\rightarrow\gamma\gamma$ and $a\rightarrow\ell^+\ell^-$ at the LHC have been performed providing significant sensitivities in large regions of unprobed parameter space \cite{Bauer:2017nlg,Bauer:2017ris,Chatrchyan:2012cg,Khachatryan:2017mnf}. These channels display reasonable sensitivity to the ALP couplings to photons and leptons for ALP masses above a few MeV. The LEP data on the decay $Z\rightarrow \gamma\gamma$ provides unprecedented access to the ALP mass range MeV to 90 GeV \cite{Jaeckel:2015jla}. It is also shown that the tri-$\gamma$ channel (with two photons originating from a prompt ALP decay) provides the possibility to constrain the mass of the ALP and its coupling to photons \cite{Mimasu:2014nea,Bauer:2017ris,Jaeckel:2015jla}. The mono-Higgs signal $pp\rightarrow a h$ is investigated in Ref. \cite{Brivio:2017ije}. The same reference also provides the prospects for bounds on the ALP coupling to fermions using the associated production of an ALP with a pair of top quarks. The ultra-peripheral heavy-ion collisions at the LHC can also be used to probe ALPs albeit these collisions are not optimized for BSM searches. The ALP is produced exclusively from the photon-photon luminosity induced by the lead ions, and decays into a pair of back-to-back photons. Utilizing this process, the ALP mass range 100 MeV to 100 GeV can be probed \cite{Knapen:2016moh}. Besides the collider experiments, the search for ALPs has also been extensively performed in proton beam-dump experiments. Long-lived ALPs with masses in the MeV-GeV range have been probed in beam-dump experiments providing new constraints on ALPs \cite{Dobrich:2019dxc}. Employing a data-driven method, the ALP-gluon coupling is carefully studied for QCD-scale ALP masses in Ref. \cite{yotam}.

In this work, we present for the first time collider searches for the production of an ALP in association with a di-jet at hadron colliders ($pp\rightarrow a+jj$) as a complementary probe to mono-jet searches. This process shows a reasonable sensitivity to the ALP coupling to gluons $c_{GG}$ and may help improve the present limits on this coupling. We simulate the detector response for the signal and background events based on the CMS detector at the LHC. A complete set of background processes is considered in our analysis and it is seen that the backgrounds are well under control. The multijet production which is the dominant background process displays properties that help to achieve an efficient signal-background discrimination using a multivariate analysis. The discriminating variables are carefully chosen so that the best discrimination is obtained. Utilizing this signal process, we set expected upper limits on the ALP-gluon coupling at the LHC (13 TeV) for light ALPs with masses $m_a=0.1,\,0.4,\,0.7$ and $1$ MeV. It is shown that the obtained limit is complementary to limits already obtained at the LHC. In addition to the constraints obtained for the LHC, we also present prospects for the HE-LHC (27 TeV) and FCC-hh (100 TeV) to gain an impression of the capability of future colliders to probe the parameter space of the present model.

This paper is structured as follows: In section \ref{sec:lagrangian}, we review the effective Lagrangian for an ALP interacting with SM gauge bosons and matter fields. In section \ref{sec:decay}, we summarize the most important decays an ALP can experience, and then we discuss the stability of ALPs in the detector. In section \ref{sec:collidersearch}, we present collider searches for the ALP production in association with two jets and set upper bounds on the ALP coupling to gluons. 

\section{Effective Lagrangian for ALPs}  
\label{sec:lagrangian}
It has long been advocated that incorporating an anomalous global $U\mathrm{(1)}$ symmetry (which is spontaneously broken) in the SM full Lagrangian provides the possibility of resolving the so-called strong CP problem by dynamically driving the parameter $\bar{\theta}$ (which controls the size of CP violation in the QCD sector) to zero \cite{Peccei:1977hh}. The pseudo Nambu-Goldstone boson associated with the spontaneously broken symmetry is called QCD axion. There is a strict relation between the mass and couplings of the axion which limits the parameter space of the axion model. Axion-like particles or ALPs are similar to QCD axions in many respects. They, however, benefit from wide-ranging masses and couplings as they avoid the stringent relation connecting the mass of the particle to its couplings. The ALP is a scalar, singlet under the SM charges and odd under the CP transformation. Neglecting ALP couplings to fermions (of any type) for the sake of simplicity, the most general bosonic effective Lagrangian up to dimension-5 operators which describes an ALP interacting with SM fields is given by~\cite{Brivio:2017ije}
\begin{eqnarray}
\begin{aligned}
	\mathscr{L}_{\mathrm{eff}} &= \mathscr{L}_{\mathrm{SM}} + \frac{1}{2} (\partial^\mu a)(\partial_\mu a) - \frac{1}{2} m_{a,0}^2 a^2\\
	&+ c_{a \Phi} \frac{i \partial^\mu a}{f_a} (\Phi^\dagger \overleftrightarrow{D}_\mu \Phi)
	-  c_{BB} \frac{a}{f_a} B_{\mu\nu} \tilde{B}^{\mu\nu}
	-  c_{WW} \frac{a}{f_a} W^{A}_{\mu\nu} \tilde{W}^{\mu\nu,A}
	-  c_{GG} \frac{a}{f_a} G^{A}_{\mu\nu} \tilde{G}^{\mu\nu,A},
\end{aligned}
\label{lag}
\end{eqnarray} 
where $a$ and $\Phi$ denote the ALP field and the Higgs boson doublet, respectively and $f_a$ is the scale associated with the breakdown of the global $U\mathrm{(1)}$ symmetry. In this equation, $G^A_{\mu\nu}$, $W^A_{\mu\nu}$ and $B_{\mu\nu}$ are the gauge field strength tensors corresponding to $SU\mathrm{(3)}_c$, $SU\mathrm{(2)}_L$ and $U\mathrm{(1)}_Y$, respectively, the dual field strength tensors $\tilde{X}^{\mu\nu}$ are defined as $\tilde{X}^{\mu\nu} \equiv \frac{1}{2} \epsilon^{\mu \nu \rho \sigma} X_{\rho \sigma}$, where $\epsilon^{\mu \nu \rho \sigma} $ is the Levi-Civita symbol, and $\Phi^\dagger \overleftrightarrow{D}_\mu \Phi \equiv \Phi^\dagger D_\mu \Phi - (D_\mu \Phi)^\dagger \Phi$. The Lagrangian of Eq. \ref{lag} is invariant under the SM gauge transformations. In this Lagrangian, promoting the $\bar{\theta}$ parameter to a dynamical variable as proposed in the axion model \cite{Peccei:1977hh}, the CP-violating $\bar{\theta}$-term in the QCD sector of the SM Lagrangian $\mathscr{L}_{\mathrm{SM}}$ has been replaced with the CP-conserving $aG\tilde{G}$ operator (the last operator). As a result, the $\bar{\theta}$-induced CP violation in the QCD sector vanishes and the strong CP problem no longer exists. 

Under a $U\mathrm{(1)}$ transformation, the ALP field undergoes the translation $a(x)\rightarrow a(x)+\alpha$, with $\alpha$ constant, implying that if the effective Lagrangian is to be $U\mathrm{(1)}$ invariant, it must also respect the symmetry under the ALP field translation (shift symmetry). The ALP field must thus only show up with a derivative in the Lagrangian. This, however, is not the case because of the chiral anomaly in this model. In the $\mathscr{L}_{\mathrm{eff}}$ of Eq.~\ref{lag}, the last operator, which couples the ALP field directly to the gluon density $G\tilde{G}$, induces a modification to the action when the ALP field translates. The gluon density is a total divergence $G^{a}_{\mu\nu} \tilde{G}^{a\mu\nu}\equiv \partial_\mu K^\mu$, and thus the $aG\tilde{G}$ operator induces the total divergence contribution $\frac{\alpha}{f_a}\partial_\mu K^\mu$ to $\delta\mathscr{L}_{\mathrm{eff}}$ under the $a$ field translation. However, the integral of this contribution doesn't vanish because of the QCD vacuum structure and instanton effects. The action is thus altered albeit a discrete version of the shift symmetry is preserved. In addition to the chiral anomaly, the shift symmetry is also affected by the explicit symmetry breaking ALP mass term present in the Lagrangian. The shift symmetry is softly broken and the ALP acquires more mass than is obtained from non-perturbative dynamics as a result of the ALP mass term. In the case of the original QCD axion~\cite{Peccei:1977hh}, where no mass term is present in the Lagrangian, the axion receives its mass solely via spontaneous breaking of the global $U\mathrm{(1)_{PQ}}$ symmetry. Therefore, the axion mass is stringently related to the characteristic scale of global symmetry breaking $f_a$ via
\begin{eqnarray}
m_a=m_{a,\,\textrm{dyn}}={\left[\left<\frac{\partial^2 V_\mathrm{eff}}{\partial a^2}\right>\right]}^{\frac{1}{2}} \propto \frac{1}{f_a},
\label{ma}
\end{eqnarray}
where $V_\mathrm{eff}$ represents the effective potential for the axion field provided by the $aG\tilde G$ operator in the Lagrangian. The ALP mass, however, doesn't suffer from such a strict relation by virtue of the presence of the soft symmetry breaking mass term $m_{a,0}^2$ in $\mathscr{L}_{\mathrm{eff}}$, which adds a mass to the dynamically generated mass: $m_a^2=m_{a,0}^2+m_{a,\,\textrm{dyn}}^2$, making $m_a$ and $f_a$ two independent parameters. This freedom opens up a wide range of possibilities and thus a rich phenomenology of ALPs. 

After electroweak symmetry breaking (EWSB), a product term of the form $Z_\mu \partial^\mu a$ which stems from the $c_{a \Phi} \frac{i \partial^\mu a}{f_a} (\Phi^\dagger\overleftrightarrow{D}_\mu \Phi)$ operator appears in $\mathscr{L}_{\mathrm{eff}}$. Accordingly, it is understood that besides the SM would-be Nambu-Goldstone boson, the $a$ field contributes to the longitudinal component of the $Z$ boson too. As, in this situation, the direct interpretation of the Lagrangian leads to difficulties, a field redefinition of the general form
\begin{equation}
\begin{aligned}
\Phi\to& \, e^{i\alpha_{\Phi}\, a/f_a}\Phi,\\
\psi_L\to& \, e^{i\alpha_{\psi L}\, a/f_a}\psi_L,\\
\psi_R\to& \, e^{i\alpha_{\psi R}\, a/f_a}\psi_R,
\label{redef}
\end{aligned}
\end{equation}
where $\alpha_{\Phi}$ is a real constant, $\alpha_{\psi L,R}$ are $3\times 3$ hermitian matrices in flavor space, $\psi_L = \{Q_L, L_L\}$ and $\psi_R = \{u_R, d_R, e_R\}$ is applied so as to illustrate the physical consequences of the induced $a$-$Z$ vertex. The parameters $\alpha_\Phi$ and $\alpha_{\psi L,R}$ can be conveniently chosen so that the $a$-$Z$ vertex is eliminated in favour of fermionic couplings. The $a$-$Z$ vertex is redefined away and replaced with fermionic couplings by the choice $\alpha_\Phi=c_{a \Phi}$, and tuning the parameters $\alpha_{\psi L,R}$ provides the possibility to control the structure of the induced fermionic couplings (which can be Yukawa-like, vector-axial or a combination of them)~\cite{Brivio:2017ije}. We choose to use the field redefinition of Eq.~\ref{redef} with $\alpha_\Phi = c_{a \Phi}$ and $\alpha_{\psi L,R}=0$, which, when acts on $\mathscr{L}_{\mathrm{SM}}$, yields
\begin{eqnarray} 
\begin{aligned}
	\mathscr{L}_{\mathrm{SM}} \to \,\, \mathscr{L}_{\mathrm{SM}} &+ c_{a \Phi} \left[i\left(\bar{Q}_L \mathbf{Y}_U\tilde\Phi u_R-\bar{Q}_L \mathbf{Y}_D\Phi d_R-\bar{L}_L\mathbf{Y}_E\Phi e_R\right)\frac{a}{f_a} + \text{h.c.}\right]\\
	&- c_{a \Phi} \frac{i \partial^\mu a}{f_a} (\Phi^\dagger \overleftrightarrow{D}_\mu \Phi),
\end{aligned}
\label{smredef}
\end{eqnarray} 
where $\mathbf{Y}_U$, $\mathbf{Y}_D$ and $\mathbf{Y}_E$ represent $3\times3$ matrices in flavor space containing Yukawa couplings for up-type quarks, down-type quarks and charged leptons, respectively, and where $\tilde\Phi = i\sigma^2\Phi^*$. In Eq.~\ref{smredef}, the last operator stems from the Higgs kinetic energy term in $\mathscr{L}_{\mathrm{SM}}$ while the Yukawa-like ALP-fermion operators in the first line stem from the SM Yukawa terms. The ALP-fermion operators in this equation can be conveniently written as
\begin{eqnarray}  
	c_{a \Phi} \left[i \frac{a}{f_a} \sum_{\psi=Q,L} \left(\bar{\psi}_L \mathbf{Y}_\psi \mathbf{\Phi} \sigma_3 \psi_R\right) + \text{h.c.}\right],
\label{fermionicops}
\end{eqnarray} 
where $\sigma_3$ acts on weak isospin space, $Q_R\equiv\{u_R,d_R\}$, $L_R\equiv\{0,e_R\}$ and the block matrices $\mathbf{\Phi}$ and $\mathbf{Y}_\psi$ are given by
\begin{eqnarray}
	\mathbf{\Phi} = \textrm{diag}(\tilde\Phi, \Phi), \ \ \ \ 
	\mathbf{Y}_Q \equiv \textrm{diag}(\mathbf{Y}_U,\mathbf{Y}_D), \ \ \ \ 
	\mathbf{Y}_L \equiv \textrm{diag}(0,\mathbf{Y}_E).
\label{blockmatrices}
\end{eqnarray} 
Applying the field redefinition, the $c_{a \Phi} \frac{i \partial^\mu a}{f_a} (\Phi^\dagger \overleftrightarrow{D}_\mu \Phi)$ operator in $\mathscr{L}_{\mathrm{eff}}$, which induces the $a$-$Z$ vertex on EWSB, is entirely canceled out by the same operator with opposite sign in Eq.~\ref{smredef}. Moreover, the fermionic operators of Eq.~\ref{fermionicops} are added to $\mathscr{L}_{\mathrm{eff}}$. The effective Lagrangian up to dimension-5 operators is thus recast as~\cite{Brivio:2017ije}
\begin{eqnarray}
\begin{aligned}
	\mathscr{L}_{\mathrm{eff}}
	&= 
	\mathscr{L}_{\mathrm{SM}} + \frac{1}{2} (\partial^\mu a)(\partial_\mu a) - \frac{1}{2} m_{a,0}^2 a^2
	+ c_{a \Phi} \left[i \frac{a}{f_a} \sum_{\psi=Q,L} \left(\bar{\psi}_L  \mathbf{Y}_\psi \mathbf{\Phi} \sigma_3 \psi_R\right) + \text{h.c.}\right]\\
   &-  c_{BB} \frac{a}{f_a} B_{\mu\nu} \tilde{B}^{\mu\nu}
     	-  c_{WW} \frac{a}{f_a} W^{A}_{\mu\nu} \tilde{W}^{\mu\nu,A}
	-  c_{GG} \frac{a}{f_a} G^{A}_{\mu\nu} \tilde{G}^{\mu\nu,A}.
\end{aligned}
\label{lagfinal}
\end{eqnarray} 

After EWSB, the operators $aB\tilde{B}$ and $aW\tilde{W}$ in Eq.~\ref{lagfinal} induce the ALP couplings to the photon and the $Z$ boson as
\begin{eqnarray}
	\mathscr{L}_{\mathrm{eff}}
	\owns 
	-  c_{\gamma\gamma} \frac{a}{f_a} F_{\mu\nu} \tilde{F}^{\mu\nu}
	-  c_{\gamma Z} \frac{a}{f_a} F_{\mu\nu} \tilde{Z}^{\mu\nu}
	-  c_{ZZ} \frac{a}{f_a} Z_{\mu\nu} \tilde{Z}^{\mu\nu},
\label{lagFZ}
\end{eqnarray} 
where $F_{\mu\nu}$ is the electromagnetic field strength tensor, $Z_{\mu\nu}$ is the corresponding field strength tensor for the $Z$ boson and the Wilson coefficients $c_{\gamma\gamma}$, $c_{\gamma Z}$ and $c_{ZZ}$ are given by
\begin{eqnarray}
	c_{\gamma\gamma} = c_\theta^2 c_{BB} + s_\theta^2 c_{WW}, \ \ \ \ 
	c_{\gamma Z} = s_{2\theta} (c_{WW}-c_{BB}), \ \ \ \ 
	c_{ZZ} = s_\theta^2 c_{BB} + c_\theta^2 c_{WW},
\label{wilsoncoeff}
\end{eqnarray} 
with $s_\theta$ ($c_\theta$) the sine (cosine) of the weak mixing angle.

\section{ALP decay and stability at colliders} 
\label{sec:decay}
The leading interactions of the ALP described by the effective Lagrangian, Eq.~\ref{lagfinal}, govern ALP decays into pairs of SM particles. In this section, we summarize the most prominent decay modes of the ALP. To keep the discussion pertinent to the scope of the paper, only decays relevant to MeV-scale ALPs are considered. In this mass region, the decays into photons, charged leptons and light hadrons are of great importance.

The di-photon decay mode plays a crucial role in many scenarios in the case of light ALPs. Particularly in the case of ALP masses below the on-shell electron pair production threshold $m_a<2m_e \approx 1.022$~MeV, as in the present work, where it is the only possible decay mode. At leading order, the decay rate of an ALP into a photon pair in terms of the relevant Wilson coefficient is given by
\begin{equation} 
\Gamma(a \rightarrow \gamma\gamma) = \frac{m_a^3}{4\pi} \Bigl(\frac{c_{\gamma\gamma}}{f_a}\Bigr)^2.
\label{widtha2gaga}
\end{equation} 
Beyond the tree-level, other Wilson coefficients in the effective Lagrangian are also involved in the decay rate. For the decay rate of Eq. \ref{widtha2gaga} to include higher-order corrections, it proves convenient to substitute $c_{\gamma\gamma}$ in this equation with the effective Wilson coefficient $c_{\gamma\gamma}^\mathrm{eff}$ in which the loop-induced effects are encoded~\cite{Bauer:2017ris}. At the one-loop level, fermion loops and electroweak loops involving all the Wilson coefficients in the Lagrangian Eq. \ref{lagfinal} except for $c_{GG}$ contribute to the corrections. Corrections due to the ALP-gluon coupling first appear at two-loop level and can be significant because of a logarithmic dependence on the global symmetry breaking scale. Even if $c_{\gamma\gamma}$ vanishes, contributions from other Wilson coefficients to $c_{\gamma\gamma}^\mathrm{eff}$ can result in a non-zero decay rate. The effective coefficient $c_{\gamma\gamma}^\mathrm{eff}$ is also dependent on the ALP mass. This dependence is, however, so mild that the decay rate varies as $m_a^3$ to an excellent approximation (see Eq. \ref{widtha2gaga}). As $m_a$ decreases, the decay rate decreases substantially leading to so large lifetimes for very light ALPs with $m_a<2m_e$.

The leptonic decay $a\rightarrow \ell^+\ell^-$ (with $\ell=e,\mu,\tau$) becomes available for ALP masses larger than $2m_\ell$. Depending on the Wilson coefficients, leptonic decays can be significant or not. To gain some insight, it's worth mentioning that in the case where the Wilson coefficients share the same magnitudes and the ALP mass does not exceed the pion mass $m_\pi$ (assuming the ALP couples to quarks and gluons as well), the leptonic modes are dominant in the vicinity of pair production thresholds $2m_\ell$~\cite{Bauer:2017ris}. In other regions of parameter space, however, the ALP decays predominantly into photons. 
 
The hadronic decays of the ALP start to appear when the ALP mass exceeds $m_\pi$. Such processes begin with the ALP decay into colored particles at the partonic level, i.e. $a\rightarrow gg$ and $a\rightarrow q\bar{q}$. In the mass range $m_a<1$ GeV, the triple pion decay modes $a\rightarrow 3\pi^0$ and $a\rightarrow \pi^0 \pi^- \pi^+$ are the dominant hadronic modes. Other decay modes possible in this range are substantially suppressed. For instance, the modes $a\rightarrow \pi\pi\gamma$, $a\rightarrow \pi^0\gamma\gamma$ and $a\rightarrow \pi^0 e^- e^+$, although allowed, are substantially suppressed by phase space and powers of the fine-structure constant~\cite{Bauer:2017ris}.    
  
After the brief preliminary discussion on the ALP decay modes, we now turn to the important subject of stability of ALPs at colliders. If the total decay width of the ALP $\Gamma_a$ is sufficiently large, the ALP undergoes a prompt or displaced decay inside the detector after production. It can be then reconstructed by its decay products. On the other hand, in the case where the total decay width is small enough, the ALP escapes the detector before its decay and manifests itself as missing transverse energy $\slashed{E}_T$. Apart from the total decay width, the Lorentz boost factor $\gamma$ plays an important role in detectability of ALPs. It, however, depends on the employed experimental setup. The average decay length of the ALP perpendicular to the beam axis $L_a^\perp$ and the probability that the ALP decays inside the detector $P_a^{\mathrm{\,det}}$ are given by
\begin{equation} 
L_a^\perp(\theta) = \gamma\beta\tau\sin\theta = \frac{\sqrt{\gamma^2-1}}{\Gamma_a}\sin\theta \equiv L_a\sin\theta, \ \ \ P_a^{\mathrm{\,det}}=1-e^{-L_{\mathrm{det}}/L_a^\perp(\theta)}, 
\label{adecaylengthprob}
\end{equation} 
where $\theta$ is the angle of the ALP's flight direction with respect to the beam axis, $\beta$ is the ALP speed, $\tau$ is the proper lifetime and $L_{\mathrm{det}}$ is the transverse distance of the calorimeter from the collision point. To get a sense of the size of ALP decay length, we find an approximate lower limit of the ALP decay length and compare it with the size of LHC detectors ($\sim 10\,\mathrm{m}$). For an ALP of mass 1 MeV, which will be considered later in this work, the di-photon decay is the most significant decay mode affecting our evaluation. The ALP decay length in the laboratory frame can be therefore estimated as
\begin{equation} 
L_a \approx \frac{|\vec{p}_a|}{m_a}\frac{1}{\Gamma(a\rightarrow\gamma\gamma)} = \frac{4\pi}{m_a^4}\Bigl(\frac{ f_a}{c_{\gamma\gamma}}\Bigr)^2 |\vec{p}_a|,
\label{estimate1}
\end{equation}  
where $\vec{p}_a$ is the three-momentum of the ALP. In the last step, we have used Eq.~\ref{widtha2gaga}. We assume the ALP mass to be 1 MeV. As all values above $1\times10^{-9}\,\, \mathrm{TeV^{-1}}$ have been experimentally excluded for the ratio $c_{\gamma\gamma}/f_a$ \cite{Bauer:2017nlg,Bauer:2017ris}, Eq.~\ref{estimate1} implies
\begin{equation} 
L_a \gtrsim \left(\frac{2.3\,|\vec{p}_a|}{\mathrm{GeV}}\right) 10^{21} \,\mathrm{m}.
\label{estimate2} 
\end{equation}
The ALP momentum $|\vec{p}_a|$, which is closely related to the missing transverse energy of the event, varies with the experiment conditions. It depends on the center-of-mass energy of the collider, the imposed minimum $\slashed{E}_T$ cut, technical specifications of the calorimetry system, etc. Assuming that the order of magnitude of $|\vec{p}_a|$ is roughly $>\mathcal{O}(100)$ GeV, one finds $L_a \gtrsim 2.3\times10^{23}\,\mathrm{m}$ according to Eq.~\ref{estimate2}. This lower limit implies that, on average, an ALP of mass 1 MeV travels a distance substantially larger than the LHC detectors radius before it decays. Such ALPs are, therefore, very likely to appear as invisible particles and the missing transverse energy reveals their existence. Apart from the di-photon decay mode, the modes $a\rightarrow\nu\bar{\nu}\nu\bar{\nu}$ and $a\rightarrow\gamma\nu\bar{\nu}$ are also available for the ALP mass considered here. However, they do not alter our conclusion. The $a\rightarrow\nu\bar{\nu}\nu\bar{\nu}$ decay mode cannot be detected at the LHC and thus an ALP decaying through this channel appears invisible even if it decays inside the detector. The $a\rightarrow\gamma\nu\bar{\nu}$ mode, which could be mediated by the $a$-$Z$-$\gamma$ interaction (see Eq.~\ref{lagFZ}), would not affect the lower limit~\ref{estimate2} significantly if it was also taken into account in Eq.~\ref{estimate1} in addition to the di-photon decay mode. The reason is that the upper limit on $\Gamma(a\rightarrow\gamma\nu\bar{\nu})$ (which is inferred from the experimental constraint on the Wilson coefficient $c_{\gamma Z}$~\cite{Brivio:2017ije}) is many orders of magnitude smaller than that of $\Gamma(a\rightarrow\gamma\gamma)$ and thus the total decay width can be safely restricted to the two-photon decay width. For ALPs  lighter than 1 MeV, the ALP decay length is even larger as according to Eq. \ref{estimate1} the decay length is inversely related to $m_a^4$. Lighter ALPs have therefore more tendency to stay invisible at the detector. The change in the probability that the ALP decays inside the detector as the ALP mass decreases is however not significant due to the smallness of the detector size $L_{\mathrm{det}}$ when compared with the ALP decay length for ALP masses in this range (see Eq. \ref{adecaylengthprob}).

It can be concluded that a large fraction of ALPs under consideration in this work are stable at the detector and manifest themselves as missing energy. This leads to large missing energies compared with typical missing energies in SM processes. The ALPs which decay inside the detector produce a small effect which will be also considered in our analysis.

\section{Associated production of an ALP with two jets} 
\label{sec:collidersearch}
The ALP can be produced through various processes at a proton-proton collider. In this work, we consider the ALP production in association with two jets. Studying this process, we set expected upper bounds on the ALP-gluon coupling $c_{GG}$ at the LHC for ALPs with masses $m_a=0.1,\,0.4,\,0.7$ and $1$ MeV described by the effective Lagrangian Eq. \ref{lagfinal}. We implement this Lagrangian into \texttt{FeynRules} \cite{Alloul:2013bka} and pass the generated Universal FeynRules Output (UFO) \cite{Degrande:2011ua}\footnote{\url{http://feynrules.irmp.ucl.ac.be/attachment/wiki/ALPsEFT/ALP_linear_UFO.tar.gz}} model, which is based on the study in Ref.~\cite{Brivio:2017ije}, to \texttt{MadGraph5\_aMC@NLO}~\cite{Alwall:2011uj} for event generation and cross section computation. In order to have a correct treatment for signal generation, the ALP is generated in association with 1 parton and 2 partons, and the resulting events are merged using the MLM merging algorithm \cite{Mangano:2006rw}. In what follows in this section, we provide a detailed analysis of the assumed production process as well as the resulting constraints. We also present prospects for the HE-LHC (27 TeV) and the FCC-hh (100 TeV) and compare the results.

\subsection{Signal and background processes} 
\label{subsec:signalBG}
\begin{figure}[t]
\centering
    \begin{subfigure}[b]{0.67\textwidth} 
    \centering
    \includegraphics[width=\textwidth]{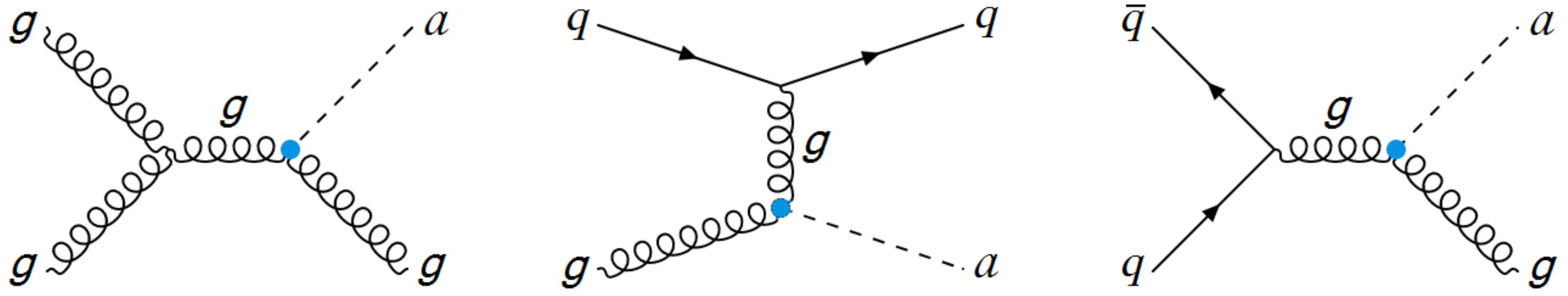}
    \end{subfigure} 
\par\bigskip
    \begin{subfigure}[b]{0.67\textwidth} 
    \centering
    \includegraphics[width=\textwidth]{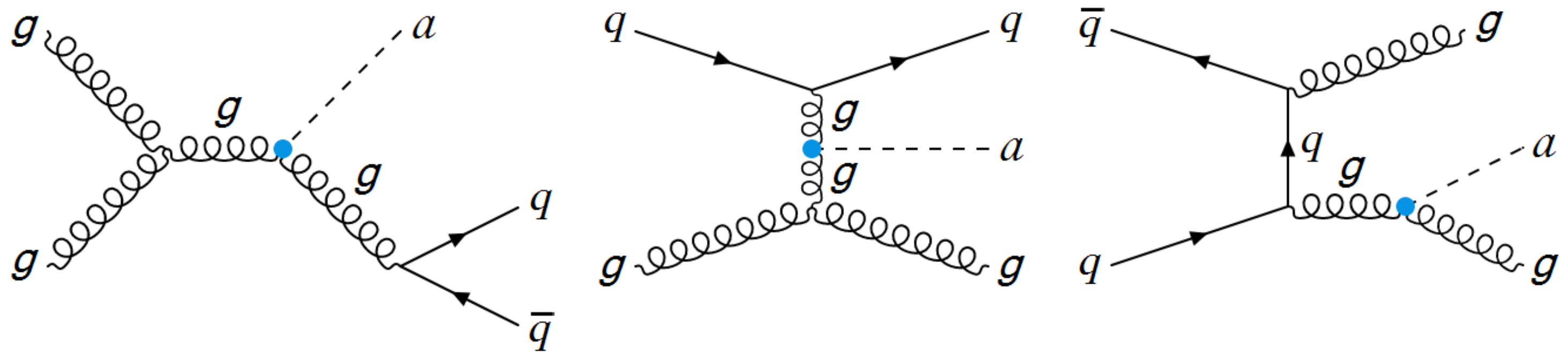}
    \end{subfigure}
  \caption{Representative Feynman diagrams relevant to the ALP production in association with up to two jets in proton-proton collisions involving only the ALP-gluon coupling.}
\label{feynmanDiagrams}
\end{figure}

In this work, the associated production of an ALP and a di-jet ($pp\rightarrow a+jj$) is assumed as the signal process. For the signal generation, it is assumed that the ALP is produced in association with one jet ($a+j$) or two jets ($a+jj$) as these processes are closely related to each other. The similarity of these processes (from now on referred as mono-jet and di-jet processes) may cause event double counting during the event generation. To remove redundant events from the generated sample, the MLM merging scheme (described in section \ref{subsec:eventgeneration}) is used in this analysis. As our goal in this study is to examine the di-jet process, we select events from the merged sample in such a way that the di-jet process has the absolute majority. The mono-jet and di-jet processes display sensitivity to the ALP-gluon and ALP-fermion couplings, i.e. $c_{GG}$ and $c_{a \Phi}$. Using \texttt{MadGraph5\_aMC@NLO} interfaced with \texttt{Pythia 8.2.43}~\cite{Sjostrand:2006za} to compute the cross section and considering the effect of one operator at a time, the leading order cross section of the ALP production in association with up to two jets ($pp\rightarrow a+j/jj$) at the center-of-mass energy of 13 TeV (as an example) can be expressed as
\begin{equation} 
\begin{split}
 \sigma^{\,a+j/jj}\,(c_{GG}/f_a)& = 4.65\times10^5 \,\Bigl(\frac{c_{GG}}{f_a}\Bigr)^2 \,\text{pb}, \,\,\,\,\,\sigma^{\,a+j/jj}\,(c_{a\Phi}/f_a) = 0.225\,\Bigl(\frac{c_{a\Phi}}{f_a}\Bigr)^2 \,\text{pb},
\end{split}
\label{xsec1} 
\end{equation}
where the unit of $f_a$ is TeV. These results have been obtained by generating a merged $a+j$ and $a+jj$ sample. In the generation, the NNPDF23 \cite{Ball:2012cx} is used as the parton distribution function (PDF) of the proton and the ALP mass is assumed to be 1 MeV. The obtained cross sections show that the mono-jet and di-jet processes are much more sensitive to $c_{GG}$ than $c_{a\Phi}$. Although the ALP is coupled to first generation (up and down) quarks at tree-level, the contributions of these couplings to the ALP production cross section are much suppressed according to Eq. \ref{xsec1}. This can be understood by the fact that the ALP couplings to the SM fermions are proportional to the Yukawa couplings (see the Lagrangian Eq. \ref{lagfinal}). The smallness of the Yukawa couplings corresponding to the up and down quarks suppresses the contributions of the ALP-quark couplings to the ALP production and the $c_{GG}$-induced processes are thus dominant. Another reason behind this dominance is that the ALP coupling to gluons is involved in more sub-processes than the coupling to fermions in both ($a+j$ and $a+jj$) processes. Furthermore, the fact that these processes are sensitive to the gluon PDF (particularly at small $x$) significantly enhances the sensitivity to the $c_{GG}$ coupling. The gluon PDF rises steeply as $x$ decreases at small $x$. This behavior leads to a substantial enhancement of the cross section of sub-processes with one or two gluons in the initial state in which the $c_{a\Phi}$ coupling is less involved. Because of negligibility of the cross section dependence on $c_{a\Phi}$, we only concentrate on the $c_{GG}$ coupling in this work and examine the potential of the di-jet process to constrain this coupling. Fig.~\ref{feynmanDiagrams} shows the representative leading order Feynman diagrams contributing to the ALP production in association with up to two jets which only involve the ALP coupling to gluons.

In addition to the LHC, we also examine the capability of the High-Energy Large Hadron Collider (HE-LHC)~\cite{HELHC} and Future Circular Collider (FCC)~\cite{Hinchliffe:2015qma} to probe the present model. These colliders are proposed as future steps to be taken after the LHC. Ultimate center-of-mass energies and integrated luminosities of these future colliders are provided in Tab.~\ref{futurecolliders}.
\begin{table}[h]
\normalsize
    \begin{center}
         \begin{tabular}{ccc}
& $\sqrt{s}\,\,$[TeV] & $L\,\,$[$\textnormal{ab}^{-1}$] \\ \Xhline{1\arrayrulewidth}
 HE-LHC & 27 & 15 \\ 
FCC-hh & 100 & 20  \\
  \end{tabular}
\caption{Ultimate center-of-mass energies and integrated luminosities of the future colliders under consideration in this study.}
\label{futurecolliders}
  \end{center}
\end{table}

\begin{figure}[t]
  \centering
  \includegraphics[width=0.525\textwidth]{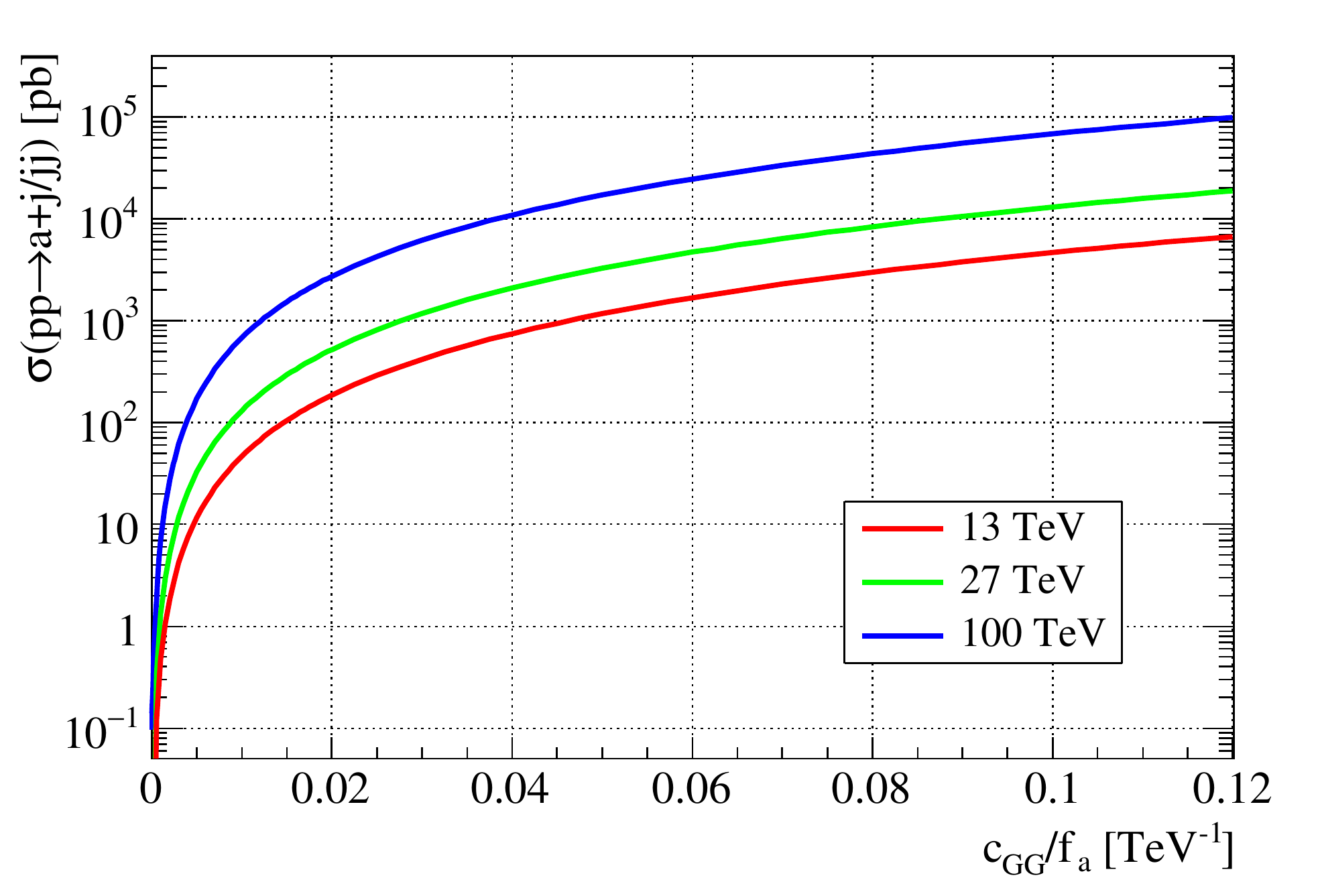}
  \caption{Cross section of the ALP production in association with up to two jets against $c_{GG}/{f_a}$ at the center-of-mass energies of 13 (red), 27 (green) and 100 (blue) TeV assuming $m_a=1$ MeV. }
\label{xsecCG}
\end{figure}

Assuming the ALP-gluon coupling to be the only non-vanishing ALP coupling and computing the leading order cross section of the $pp\rightarrow a+j/jj$ process at the center-of-mass energies of 13, 27 and 100 TeV, one finds
\begin{eqnarray} 
\begin{aligned}
 \sigma^{\,a+j/jj}\,(c_{GG}/f_a) = \lambda_{\sqrt{s}} \,\Bigl(\frac{c_{GG}}{f_a}\Bigr)^2 \,\text{pb}, \,\,\,\,\,\, \lambda_{\sqrt{s}} &=\left\{
                \begin{array}{ll}
                  4.65\times10^5 \,\,\,\,\,13 \,\mathrm{TeV} \\
                  1.30\times10^6 \,\,\,\,\, 27 \,\mathrm{TeV} \\
                  6.80\times10^6 \,\,\,\,\,100 \,\mathrm{TeV} \\
                \end{array}
              \right..
\end{aligned}
\label{xsec2} 
\end{eqnarray}
These results correspond to the ALP mass of 1 MeV and have been obtained using the same procedure and assumptions as used for obtaining the cross sections of Eq.~\ref{xsec1}. The obtained cross sections are depicted in Fig.~\ref{xsecCG}. 

Performing a scan over the ALP mass, it is observed that the cross section doesn’t display a significant mass dependence for ALP masses below few GeVs. This is not surprising though because of the smallness of the ALP mass in this range when compared with the process energy scale. With further increasing the ALP mass, the cross section will eventually decrease and become small for heavy ALPs.

The dominant SM backgrounds relevant to the assumed signal process are as follows:
\begin{itemize}
\item Multijet production.
\item Production of two massive gauge bosons $VV$ ($V=W^\pm,Z$), where the $W$ and $Z$ bosons can decay via both hadronic and leptonic decay modes.
\item $W\textnormal{+jets}$, $Z\textnormal{+jets}$ and $\gamma\textnormal{+jets}$ in all possible final states.
\item Top quark pair production ($t\bar{t}$) and single top quark production, where all the canonical channels ($s$-, $t$- and $tW$-channel) are considered in the single top production and all final states are assumed.
\end{itemize}
These background processes can produce missing transverse energy and final state jets (which resemble the signal signatures) in various ways. The sources of the missing energy in the background processes include the neutrino(s) coming from the leptonic decays of the $W^\pm$ and $Z$ bosons, $W^\pm\rightarrow\ell^\pm\nu_\ell$ ($\ell=e,\mu,\tau$) and $Z\rightarrow\nu\bar\nu$, the leptonic decays of hadrons inside the jets, uncertainties in the jet energy measurement, electronic noise of the detector, etc.

\subsection{Event generation} 
\label{subsec:eventgeneration}
Generation of the signal and background events is performed as follows. Hard (parton-level) events are generated using \texttt{MadGraph5\_aMC@NLO} and are then passed to \texttt{Pythia 8.2.43} to perform parton showering, hadronization and sequential decays of unstable particles. The output is internally passed to \texttt{Delphes 3.4.2}~\cite{deFavereau:2013fsa} to simulate the detector effect assuming an upgraded CMS detector~\cite{CMSCollaboration:2015zni,Chatrchyan:2008aa}. Jet reconstruction is performed by \texttt{FastJet}~\cite{Cacciari:2011ma} using the anti-$k_t$ algorithm~\cite{Cacciari:2008gp} with a jet cone size of 0.4. Moreover, the NNPDF23 \cite{Ball:2012cx} is used as the proton PDF.

For the signal generation, the ALP mass $m_a$ is set to 1 MeV and the ALP-gluon coupling is assumed to be the only non-vanishing coupling of the ALP. The signal sample is generated assuming that an ALP is produced in association with up to two jets. For the removal of event double counting, the MLM matching scheme \cite{Mangano:2006rw} is used. The MLM matching algorithm uses the parton shower history of events to veto events in which the parton shower generates a hard activity already generated by the matrix elements. There are two parameters used by the algorithm, xqcut and qcut, which control the jet matching quality and should be set properly to have a perfect merged sample. The xqcut variable is defined as the minimum allowed distance between partons in hard events generated in MadGraph level, and the qcut variable is defined as the jet measure cutoff (or matching scale) used by \texttt{Pythia}. The merged signal sample is generated using the values 10 GeV and 25 GeV for the xqcut and qcut parameters, respectively. Validity of the matching is checked using the differential jet rate (DJR) distributions. With this choice of parameters, the DJR distributions show a smooth transition between events with different jet multiplicities (1-jet and 2-jet events). The MLM matching is also used for the generation of multijet (merged 2-jet and 3-jet), $W\textnormal{+jets}$ (merged $W\textnormal{+\,0,1,2 jets}$) and $Z\textnormal{+jets}$ (merged $Z\textnormal{+\,0,1,2 jets}$) backgrounds. The generation of the multijet sample is performed using the values of 10 GeV and 20 GeV for xqcut and qcut, respectively. For the $W\textnormal{+jets}$ and $Z\textnormal{+jets}$ generations, xqcut and qcut are set to 10 GeV and 15 GeV, respectively. 

According to the discussion of section~\ref{sec:decay}, a very small fraction of ALPs produced in proton-proton collisions decay inside the detector and thus don't contribute to the missing transverse energy (except for those rare ALPs with invisible particles in their decay final states, e.g. the ALPs experiencing the decay mode $a\rightarrow\nu\bar\nu\nu\bar\nu$). Using the probability that an ALP decays inside the detector $P_a^{\mathrm{\,det}}$ which is defined in Eq.~\ref{adecaylengthprob}, we take this effect into account event-by-event.

\subsection{Validity of the effective Lagrangian} 
\label{subsec:validity}
Validity of the effective Lagrangian Eq. 7 can be ensured by requiring the suppression scale, $f_a$ in the present model, to be much larger than the typical energy scale $\sqrt{\hat{s}}$ of the processes under consideration. The criterion $\sqrt{\hat{s}}<f_a$ should be therefore imposed on the events. However, the processes under study have an undetectable ALP in their final states and thus $\sqrt{\hat{s}}$ cannot be experimentally measured. In this case, validity of the effective field theory may be naively ensured by requiring $2\slashed{E}_T^{max}<f_a$, where $\slashed{E}_T^{max}$ is the highest missing transverse energy data bin in the analysis. Considering the correlation between $\sqrt{\hat{s}}$ and $\slashed{E}_T$, the strict validity criterion $\sqrt{\hat{s}}<f_a$ can be implemented by discarding the fraction of events in each bin for which $\sqrt{\hat{s}}>2\slashed{E}_T^{max}$. In this analysis, we make use of this procedure and apply the strict validity criterion to ensure the validity of the effective description.

\subsection{Event selection and signal-background discrimination} 
\label{subsec:discrimination}
As for the high level trigger (HLT), events are required to satisfy the conditions $\slashed{E}_T>60$ GeV and $p_T^{\, j}>180$ GeV \cite{Virdee:1043242} where $p_T^{\, j}$ is the transverse momentum of the hardest (with the highest transverse momentum) jet with the pseudorapidity threshold $\vert \eta_j \vert < 2.5$. Events are required to have 2 or 3 reconstructed jets with $p_T>30$ GeV and $\vert \eta \vert < 2.5$. The majority of the signal events selected (from the merged signal sample) by this condition belong to the $a+jj$ production process. Events with identified isolated photons or leptons (electrons and muons) with $p_T>10$ GeV and $\vert \eta \vert < 2.5$ are vetoed. Isolated objects are selected with the help of the isolation variable $I_{rel}$ as defined in Ref. \cite{deFavereau:2013fsa}. $I_{rel}$ is required to be $<0.12$ for photons and electrons, and $<0.25$ for muons. Imposing all the conditions, event preselection efficiencies presented in Tab.~\ref{eff} are obtained for signal and background processes. We observe a sizable suppression of the overwhelming multijet and $\gamma\textnormal{+jets}$ backgrounds as a result of the preselection cuts. It is also observed that the efficiencies grow with the center-of-mass energy which is expected as higher center-of-mass energies produce harder jets and larger missing transverse energies.

\begin{figure}[!h]
\begin{center}
\includegraphics[width=0.48\textwidth]{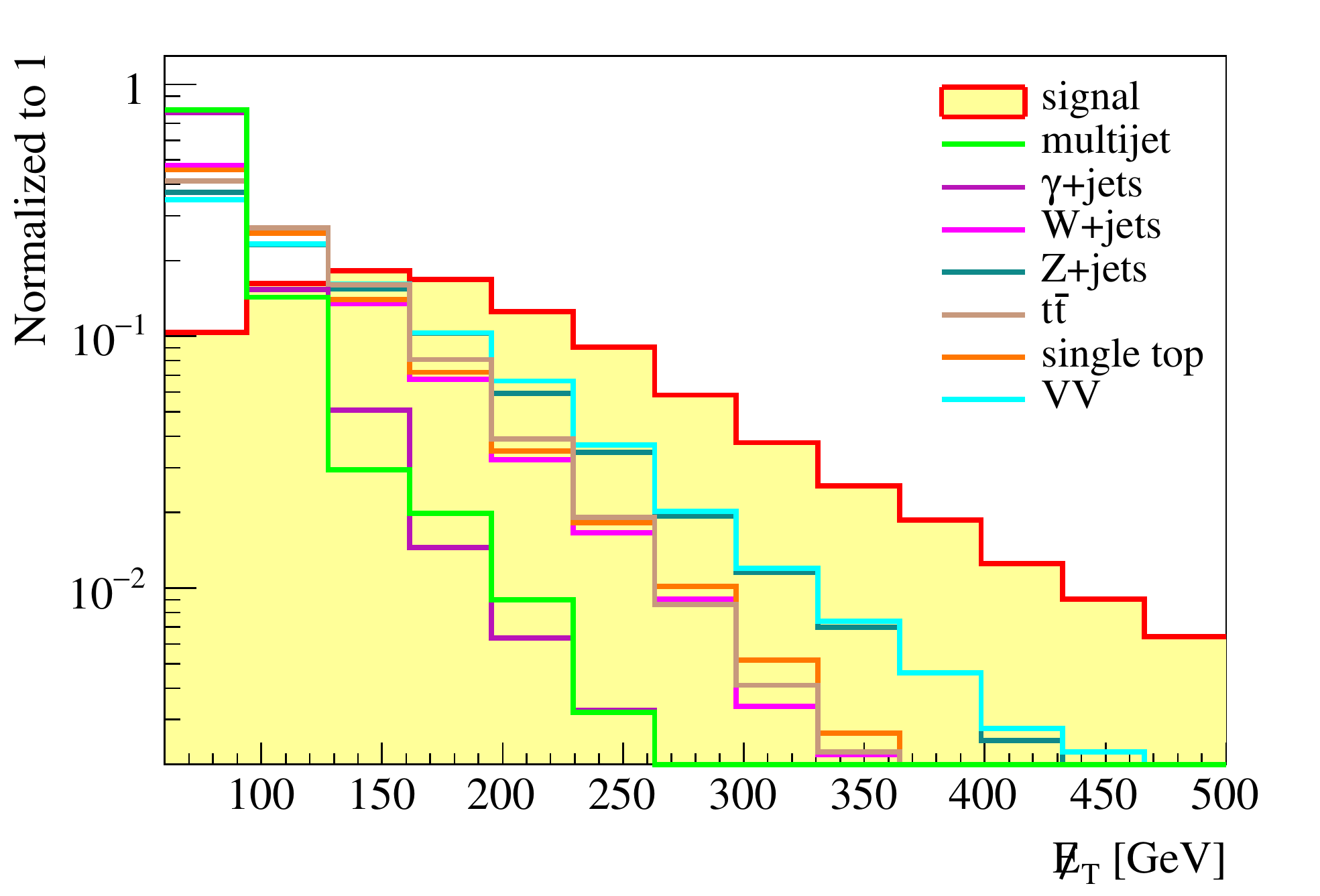}
\includegraphics[width=0.48\textwidth]{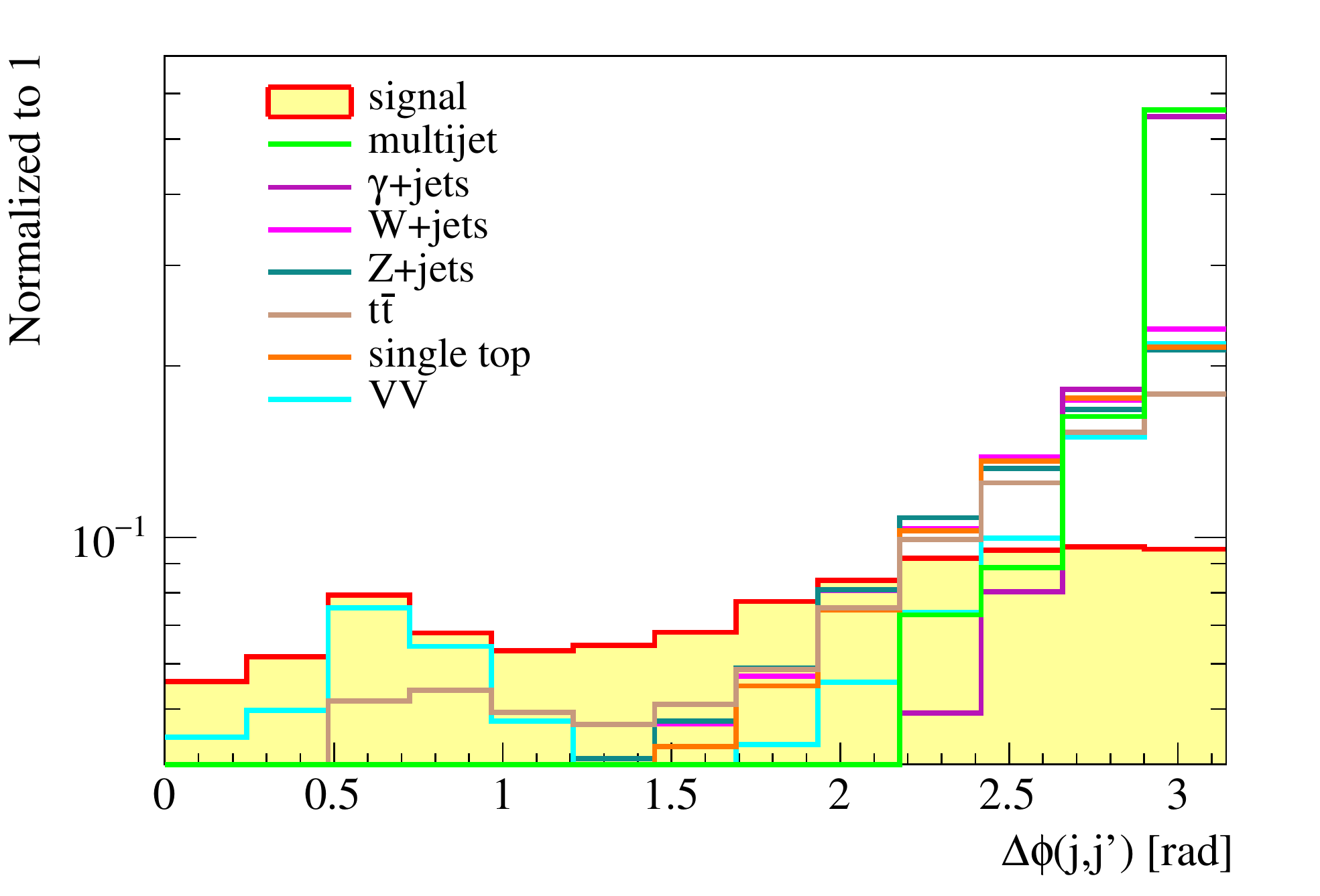} \\
\includegraphics[width=0.48\textwidth]{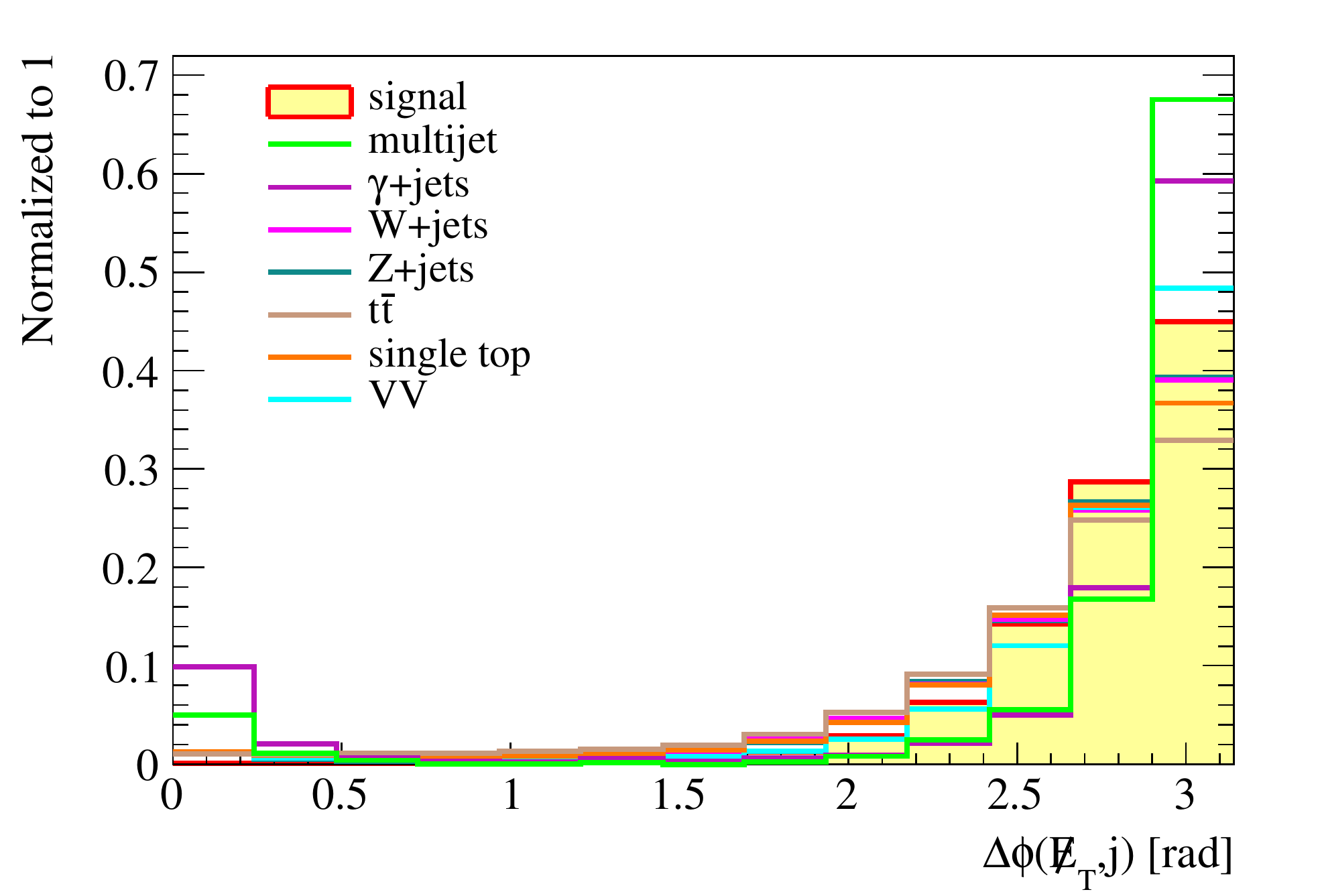} 
\includegraphics[width=0.48\textwidth]{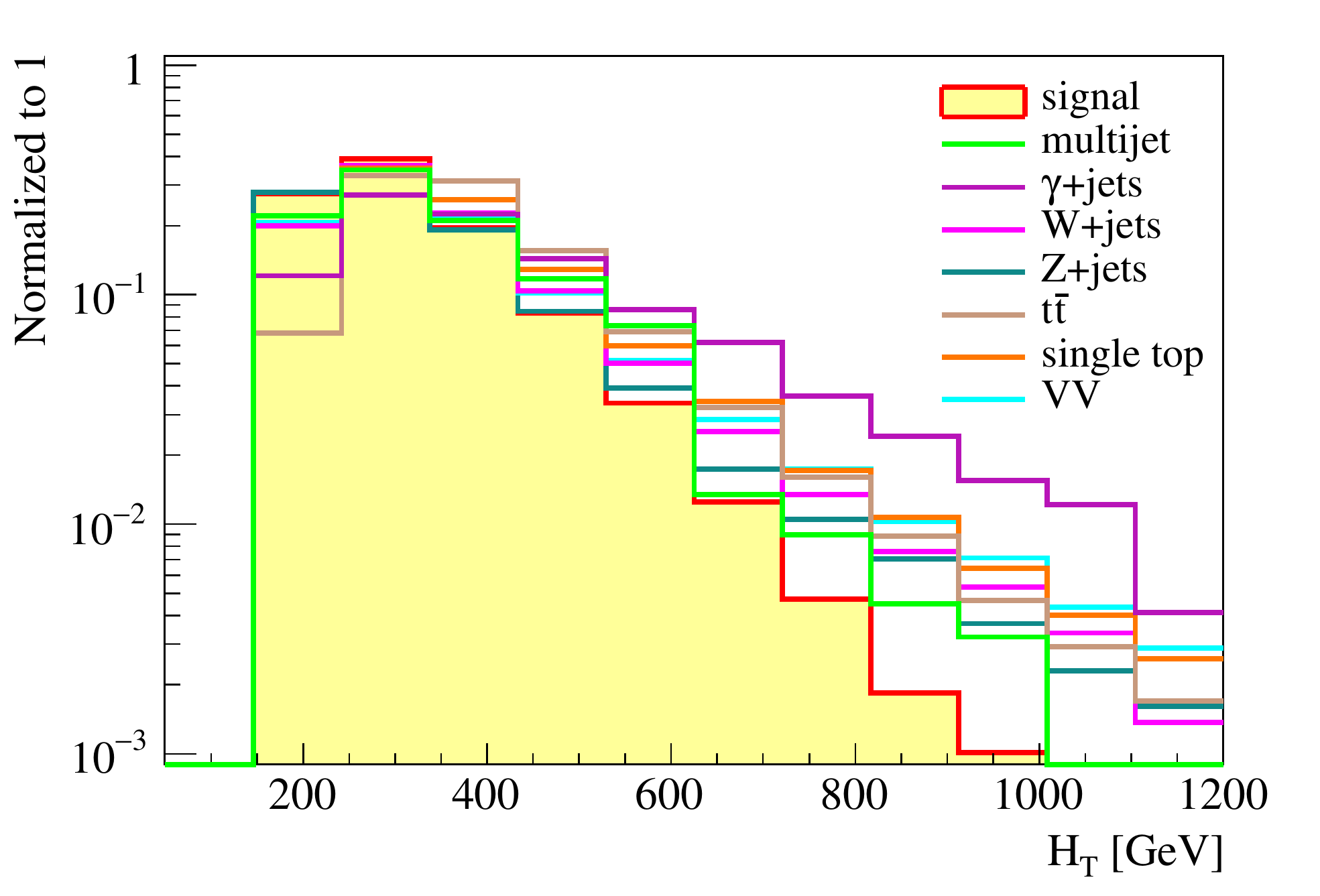} \\
\includegraphics[width=0.48\textwidth]{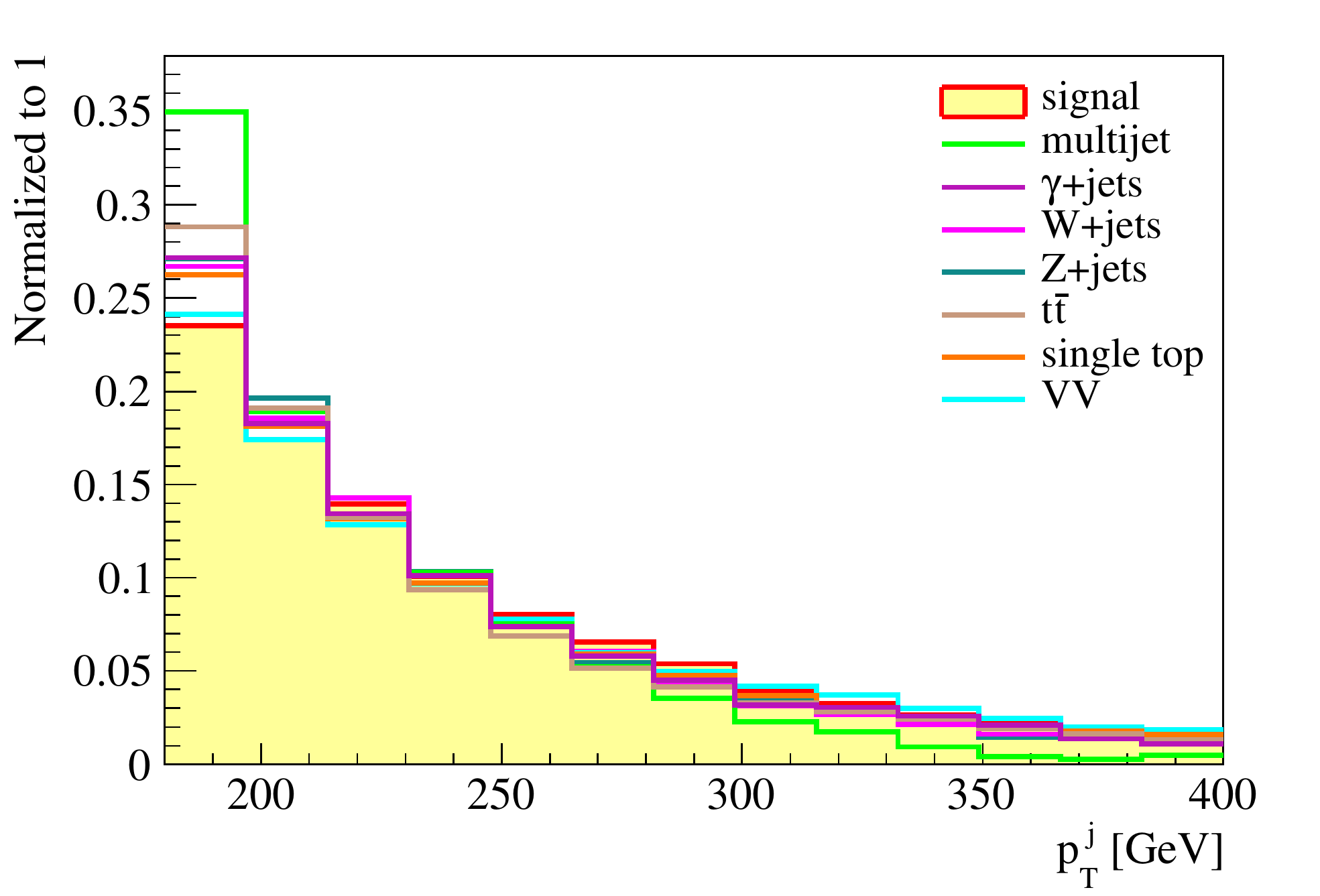} 
\includegraphics[width=0.48\textwidth]{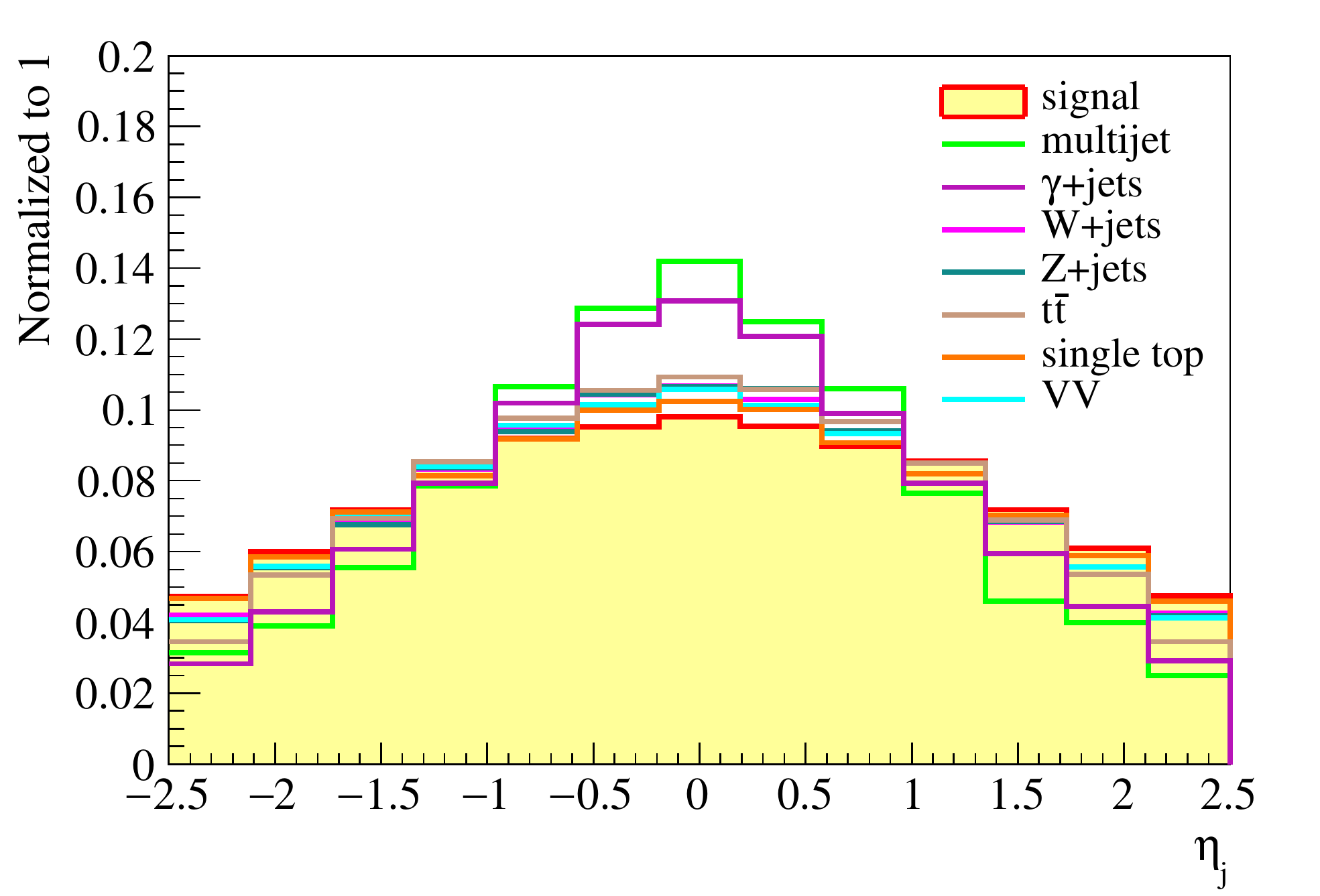} 
\caption{Distributions of some of the discriminating variables used as input to the BDT algorithm obtained at the center-of-mass energy of 13 TeV assuming $m_a=1$ MeV.}
\label{divariables}
\end{center}
\end{figure}

To better discriminate between the signal and background events, we employ a multivariate technique~\cite{Hocker:2007ht} in this analysis. The preselected events are analyzed exploiting the Boosted Decision Trees (BDT) algorithm. The BDT is fed with a proper set of discriminating variables needed to perform the training process. The discriminating variables used in this analysis are:
\begin{itemize}
    \item Missing transverse energy $\slashed{E}_T$.
    \item Scalar transverse energy sum $H_T$ which is defined by $H_T=\sum\limits_{i}|\vec{p}_T(i)|$, where the index $i$ runs over all particles reconstructed in the detector.    
    \item Azimuthal separation between the missing transverse energy and the hardest jet $\Delta\phi(\slashed{E}_T,j)$.
\item Azimuthal separation between the two hardest jets $\Delta\phi(j,j^\prime)$.
     \item Transverse momentum and pseudorapidity of the hardest jet $p_T^{\, j}$, $\eta_{j}$.
     \item Transverse momentum and pseudorapidity of the second hardest jet $p_T^{\, j^\prime}$, $\eta_{j^\prime}$.
\end{itemize}
Distributions of some of the discriminating variables obtained from the preselected events at the center-of-mass energy of 13 TeV (as an example) are presented in Fig. \ref{divariables}.

The missing transverse energy plays a significant role in separation of signal and background events. As shown in section \ref{sec:decay}, a large fraction of produced ALPs in $pp$ collisions cover a distance many orders of magnitude larger than the detector size before they decay. As a result, a missing transverse energy due to the undetectable ALPs is expected in the final state. The signal missing transverse energy is therefore expected to be significantly larger than that of the background processes since the source of missing energy in the background processes is limited to final state neutrinos. 

In addition to the missing transverse energy, the azimuthal separation between the two hardest jets is also of great importance in discriminating signal from backgrounds. Jets in di-jet events, which constitute the majority of events in the multijet background (the dominant background process), are likely to be back-to-back distributed in the $\phi$-plane because of the momentum conservation. Their azimuthal separation is therefore expected to be close to $\pi$ radians. This is, however, not the case in signal events as the signal process has a three-body final state with jets distributed randomly in the azimuth plane. As a result, the distributions of $\Delta\phi(j,j^\prime)$ for signal and multijet events show a clear contrast and have a high discriminating power.
\begin {table}[t]
\centering
         \begin{tabular}{ cccccccccc } 
  &  & signal & $\gamma\textnormal{+jets}$ & $W\textnormal{+jets}$ & $Z\textnormal{+jets}$ & $t\bar{t}$ & single top & $VV$ & multijet \parbox{0pt}{\rule{0pt}{1ex+\baselineskip}}\\ \Xhline{1\arrayrulewidth} 
   &   \cellcolor{blizzardblue}{13 TeV} & 0.0056 & 7.0e-05 & 0.0024 & 0.0036 & 0.0057 & 0.0058 & 0.0041 & 8.8e-06 \parbox{0pt}{\rule{0pt}{1ex+\baselineskip}}\\ 
\multirow{2}{*}[6.7pt] {$\epsilon_{\,\textnormal{tot}}$} & \cellcolor{blizzardblue}{27 TeV} & 0.0103 & 0.0001 & 0.0036 & 0.0050 & 0.0083 & 0.0088 & 0.0061 & 1.6e-05 \parbox{0pt}{\rule{0pt}{1ex+\baselineskip}}\\
&   \cellcolor{blizzardblue}{100 TeV} & 0.0198 & 0.0004 & 0.0061 & 0.0076 & 0.0137 & 0.0165 & 0.0093 & 3.9e-05 \parbox{0pt}{\rule{0pt}{1ex+\baselineskip}}\\ \Xhline{1\arrayrulewidth}
        \end{tabular} 
\label{effdi}
\caption {Event preselection total efficiencies ($\epsilon_{\,\textnormal{tot}}$) for the signal and background processes corresponding to the ALP mass of 1 MeV at the center-of-mass energies of 13, 27 and 100 TeV.}
\label{eff}
\end {table} 

\begin{figure}[t]
\centering
  \includegraphics[width=0.55\textwidth]{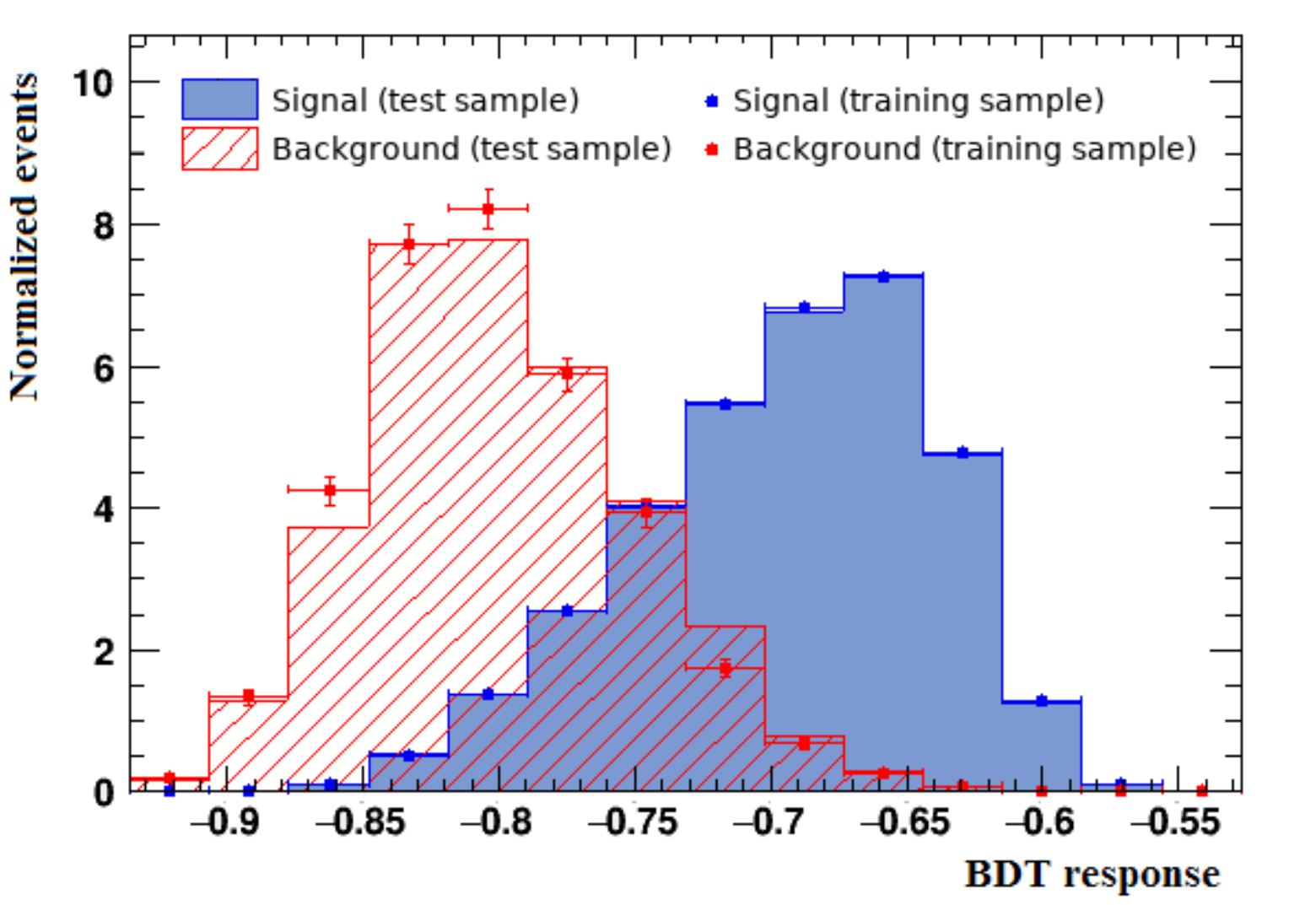}
    \caption{Distributions of the BDT response for the signal and total background at the center-of-mass energy of 13 TeV corresponding to the ALP mass of 1 MeV.}
\label{overtrainingtest}
\end{figure}

All the background processes take part in the training process according to their corresponding weights. Applying the trained model to events, the BDT classifier produces a response for each event. As an example, the distribution of the BDT response for signal and background events assuming the center-of-mass energy of 13 TeV is provided in Fig.~\ref{overtrainingtest}. As seen, the BDT performs well at separating signal events from backgrounds. This was expected because of the discriminating power the training variables provide. Particularly, $\slashed{E}_T$ and $\Delta\phi(j,j^\prime)$ variables play a significant role as discussed before. The BDT output has been examined in terms of the discriminating power using the receiver operating characteristic (ROC) curve.

\begin{figure}[t]
  \centering
  \includegraphics[width=0.57\textwidth]{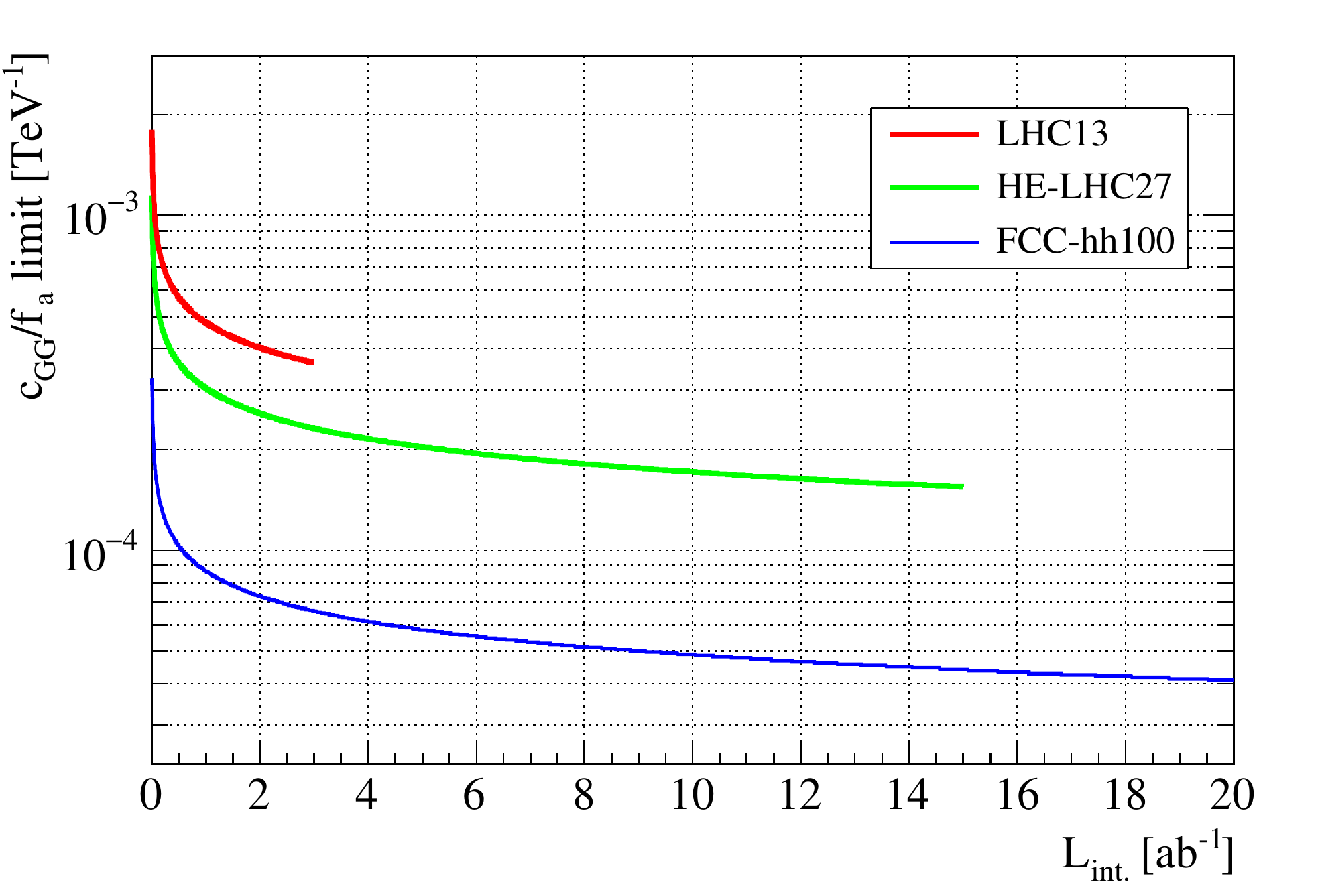}
  \caption{Expected upper limits on $c_{GG}/f_a$ at $95\%$ CL corresponding to $m_a=1$ MeV as a function of the integrated luminosity assuming the LHC 13 TeV (red), HE-LHC 27 TeV (green) and FCC-hh 100 TeV (blue).}
\label{limL}
\end{figure}
\begin{table}[t]
\normalsize
    \begin{center}
         \begin{tabular}{ccccccccc}
Process & Multijet & $\gamma$+jets & W+jets & Z+jets & $\mathrm{t\bar{t}}$ & single top & WW & ZZ \\ \Xhline{1\arrayrulewidth}
$\Delta\sigma/\sigma$ & 100\% & 5\% \cite{Aaboud:2017kff} & 10\% \cite{Sirunyan:2019dyi} & 17\% \cite{Aad:2020sle} & 4.6\% \cite{Aad:2020tmz} & 15\% \cite{Sirunyan:2018rlu} & 7\% \cite{Aaboud:2019nkz} & 6\% \cite{Sirunyan:2017zjc} \\ 
  \end{tabular}
\caption{Relative uncertainties ($\Delta\sigma/\sigma$) in the measurement of the cross sections of  different background processes.}
\label{relUncer}
  \end{center}
\end{table}
\subsection{Constraints on the ALP coupling to gluons}
\label{subsec:constraint}
\begin{figure}[t]
  \centering
  \includegraphics[width=0.57\textwidth]{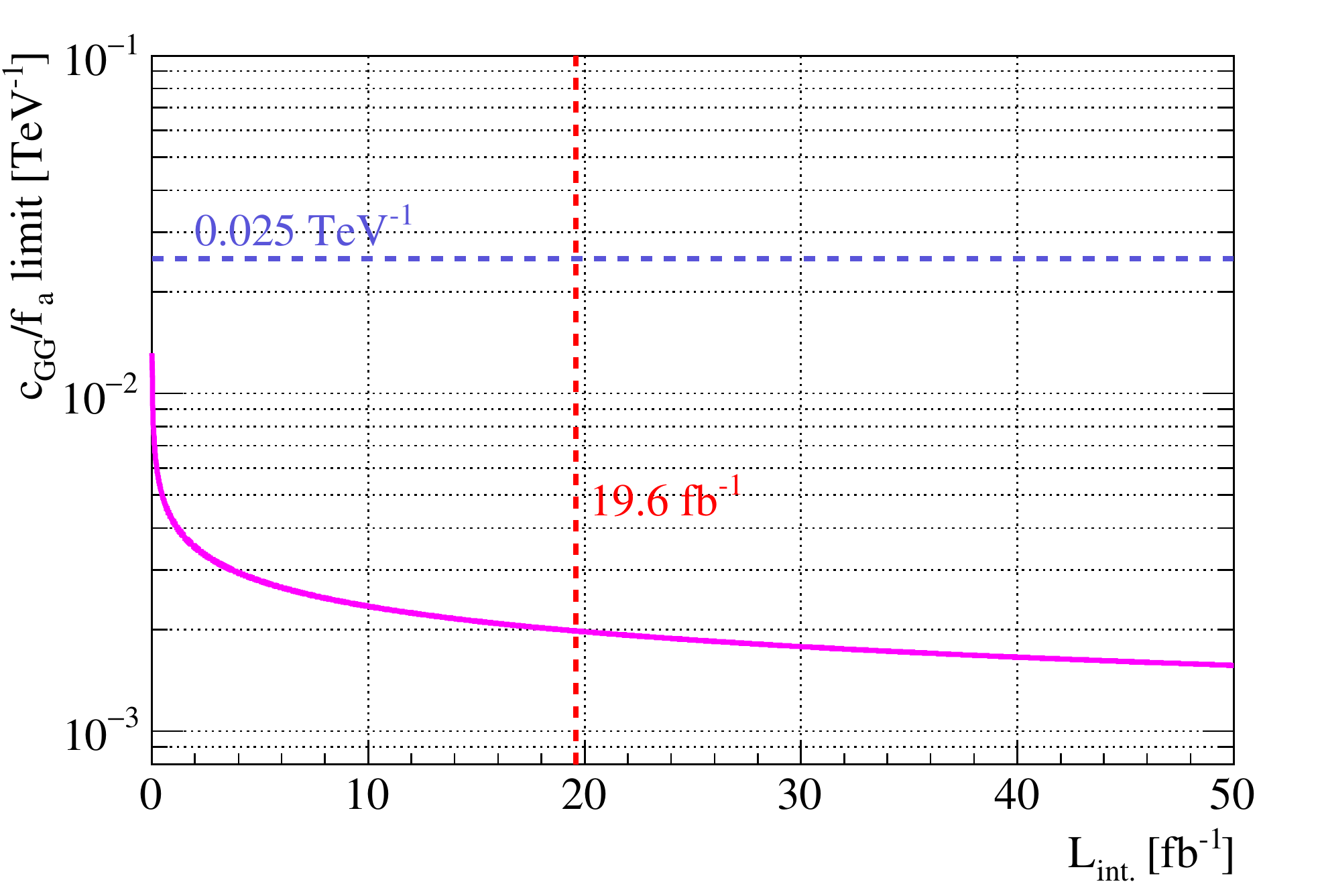}
  \caption{Expected $95\%$ CL upper limits on $c_{GG}/f_a$ as a function of the integrated luminosity assuming the LHC 8 TeV and $m_a=1$ MeV.}
\label{comparison}
\end{figure}

Using the BDT response distributions of the signal and background events, expected upper limits on the ALP coupling to gluons at the center-of-mass energies of 13, 27 and 100 TeV are computed. The obtained $95\%$ confidence level (CL) upper limits on $c_{GG}/f_a$ corresponding to the ALP mass of 1 MeV are plotted against the integrated luminosity in Fig. \ref{limL}. In this figure, different colors correspond to different center-of-mass energies. To obtain more realistic results it is necessary to take into account effects of the systematic uncertainties. To include such effects, we consider an overall uncertainty of $10\%$ on the signal selection efficiency as well as the overall uncertainties in the measurement of the cross sections of different background processes provided in Tab. \ref{relUncer}. The relative uncertainty for the multijet background is conservatively chosen to be $100\%$. To see the effect of systematic uncertainties on the ALP-gluon coupling upper limit and to better compare the limits obtained in different center-of-mass energies, upper limits (with and without including the systematic uncertainties) obtained for the LHC13, HE-LHC27 and FCC-hh100 at the integrated luminosities of 3, 15 and 20 $\mathrm{ab^{-1}}$ are presented in Tab. \ref{limits}.
\begin {table}[t]  
\centering
         \begin{tabular}{ ccccc } 
  &  & $3\,\mathrm{ab^{-1}}$ & $15\,\mathrm{ab^{-1}}$ & $20\,\mathrm{ab^{-1}}$ \parbox{0pt}{\rule{0pt}{1ex+\baselineskip}}\\ \Xhline{1\arrayrulewidth} 
   &   \cellcolor{blizzardblue}{LHC13 } & 3.6e-04 (4.5e-04) & -- & -- \parbox{0pt}{\rule{0pt}{1ex+\baselineskip}}\\ 
\multirow{2}{*}[6.7pt] {Limit on $c_{GG}/f_a$} & \cellcolor{blizzardblue}{HE-LHC27 } & 2.3e-04 (2.9e-04) & 1.5e-04 (1.9e-04) & --
  \parbox{0pt}{\rule{0pt}{1ex+\baselineskip}}\\
\multirow{2}{*}[6.7pt] {[TeV$^{-1}]$} &   \cellcolor{blizzardblue}{FCC-hh100} & 6.6e-05 (8.1e-05) & 4.4e-05 (5.4e-05) & 4.1e-05 (5.0e-05) \parbox{0pt}{\rule{0pt}{1ex+\baselineskip}}\\ \Xhline{1\arrayrulewidth}
    \end{tabular} 
\caption {$95\%$ CL expected upper limits on $c_{GG}/f_a$ obtained for the LHC (13 TeV), HE-LHC (27 TeV) and FCC-hh (100 TeV) at the integrated luminosities of 3, 15 and 20 $\mathrm{ab^{-1}}$. The limits in parentheses are obtained including the systematic uncertainties listed in Tab. \ref{relUncer} while other limits are computed without assuming the systematic uncertainties. The presented limits correspond to $m_a=1$ MeV.}
\label{limits}
\end{table} 
As can be seen, the strongest obtained constraints (including the systematic uncertainties) on $c_{GG}/f_a$ at the LHC13, HE-LHC27 and FCC-hh100 are $4.5\times10^{-4}$, $1.9\times10^{-4}$ and $5.0\times10^{-5}$ $\mathrm{TeV^{-1}}$, respectively. Mono-jet analyses using 19.6 fb$^{-1}$ of data at the LHC 8 TeV experiments result in the $95\%$ CL upper limit $0.025$ TeV$^{-1}$ on $c_{GG}/f_a$~\cite{Mimasu:2014nea}. To make a comparison between the limits obtained in this analysis and the limit derived from mono-jet events at the LHC 8 TeV, we repeat our analysis for the center-of-mass energy of 8 TeV following the same procedure as described in section \ref{sec:collidersearch}. Fig. \ref{comparison} provides the obtained $95\%$ CL upper limits on $c_{GG}/f_a$ (including the systematic uncertainties) at the center-of-mass energy of 8 TeV as a function of the integrated luminosity. As seen, at the integrated luminosity of 19.6 fb$^{-1}$, the obtained limit is more stringent than the limit from mono-jet analyses by roughly one order of magnitude. It can be concluded that the examined signal process in this study is a promising process which can be used to improve the present experimental constraint on the ALP coupling to gluons. We also compute the expected limits on $c_{GG}/f_a$ for a set of assumed ALP masses below 1 MeV at the LHC13. The $95\%$ CL upper limits for the ALP masses $0.1,\,0.4$ and $0.7$ MeV are respectively obtained to be $4.1\times10^{-4}$, $4.3\times10^{-4}$ and $4.4\times10^{-4}$ $\mathrm{TeV^{-1}}$ at the integrated luminosity of $3\,\mathrm{ab^{-1}}$. The systematic uncertainties (Tab. \ref{relUncer}) have been considered in computing the presented limits. Comparing the limits obtained at different masses, it is seen that the limit varies slightly as the ALP mass changes. This can be attributed to the fact that the cross section of the production process $pp\rightarrow a+jj$ and the probability that the ALP decays inside the detector don't change significantly as the ALP mass varies in the range $m_a\lesssim1$ MeV.

\section*{Summary and conclusions} 
\label{sec:conclusions}
Axion-like particles (ALPs) are CP-odd scalars resulting from spontaneously broken global $U(1)$ symmetries incorporated in the Standard Model. Motivated by the possibilities such particles provide to address some of the SM problems, e.g. the dark matter problem, strong CP problem, baryon asymmetry problem, neutrino mass problem, etc., we examine the potential of the production process $pp\rightarrow a+jj$ to search for light ALPs and to probe the parameter space of the model at the present and future proton-proton colliders. Light ALPs are often so feebly coupled to the SM particles that their long lifetime doesn't let them decay inside the detector. They, therefore, manifest themselves as missing energy in the final state providing an interesting signature. We present expected $95\%$ CL upper limits on the ALP coupling to gluons at the LHC (13 TeV), HE-LHC (27 TeV) and FCC-hh (100 TeV). Assuming the ALP mass $m_a=1$ MeV, the expected upper limits on $c_{GG}/f_a$ (including the systematic uncertainties) at the LHC13, HE-LHC27 and FCC-hh100 are found to be $4.5\times10^{-4}$, $1.9\times10^{-4}$ and $5.0\times10^{-5}$ $\mathrm{TeV^{-1}}$, respectively. These limits correspond to the ultimate integrated luminosities the colliders will eventually operate at according to their benchmark specifications. To compare the limits obtained in this analysis with the limit $|c_{GG}/f_a|<0.025$ TeV$^{-1}$ which has been already derived from mono-jet events at the LHC 8 TeV experiments using 19.6 fb$^{-1}$ of data, we also present expected $95\%$ CL upper limits on $c_{GG}/f_a$ at the center-of-mass energy of 8 TeV. A comparison shows that the limit obtained at the center-of-mass energy of 8 TeV and at the integrated luminosity of 19.6 fb$^{-1}$ is roughly one order of magnitude stronger than the limit derived from mono-jet analyses at the same center-of-mass energy and integrated luminosity. Therefore, the present analysis could improve the limit from mono-jet analyses at the LHC8. Results indicate that the LHC13 and the future HE-LHC27 are capable of improving the present experimental limit on $c_{GG}/f_a$ from LHC8 by roughly two orders of magnitude. Furthermore, our prospects for bounds at FCC-hh100 indicate that this collider is able to improve the present experimental limit by roughly three orders of magnitude. Besides the limits obtained for 1 MeV ALPs, the expected upper limit on $c_{GG}/f_a$ has also been computed for some ALP masses below 1 MeV at the LHC13. Including the systematic uncertainties, the $95\%$ CL expected limits at the integrated luminosity of $3\,\mathrm{ab^{-1}}$ obtained assuming $m_a=0.1,\,0.4$ and $0.7$ MeV are $4.1\times10^{-4}$, $4.3\times10^{-4}$ and $4.4\times10^{-4}$ TeV$^{-1}$, respectively. As seen, the limit varies slightly as the ALP mass decreases. This is in accordance with our expectations since the ALP production cross section and the probability that the ALP decays inside the detector are not significantly sensitive to the ALP mass for masses in this range. It can be concluded that the present analysis can serve experimentalists well in probing the light ALP physics since the ALP-gluon coupling is reachable through this analysis in a significant portion of the parameter space. 

\section*{Acknowledgments}
The authors gratefully thank Yotam Soreq for providing data of  Ref. \cite{yotam}. 

\RaggedRight 

\end{document}